\documentclass[twocolumn,prb,showpacs,multicol,amsmath,amssymb]{revtex4}
\usepackage[dvips]{graphicx}
\usepackage{graphicx}
\usepackage{dcolumn}
\usepackage{bm}
\usepackage{graphics}
\usepackage{epsfig,color}

\newcommand{\be}{\begin{equation}}
\newcommand{\ee}{\end{equation}}
\newcommand{\bea}{\begin{eqnarray}}
\newcommand{\eea}{\end{eqnarray}}

\begin{document}
\title{Induced effects of the Dzyaloshinskii-Moriya interaction on the thermal entanglement in spin-1/2 Heisenberg chains}
\author{E. Mehran$^1$, S. Mahdavifar$^{1}$, R. Jafari$^2$}
\affiliation{ $^{1}$Department of Physics, University of
Guilan, 41335-1914, Rasht, Iran}
\affiliation{ $^{2}$Research Department, Nanosolar System Company (NSS), Zanjan 45158-65911, Iran}
\date{\today}
\begin{abstract}
  The nearest neighbor spins in the one-dimensional (1D) spin-1/2 XX model with added Dzyaloshinskii-Moriya (DM) interaction are entangled at zero temperature. In the presence of a transverse magnetic field (TF) they remain entangled up to a quantum critical field  $h_c$. Using the fermionization technique, we have studied the mutual effect of the DM interaction and TF on the thermal entanglement (TE) in this model. The critical temperature, where the entanglement disappears is specified. It is found that the TE at a finite temperature neighborhood of the quantum critical field shows a scaling behavior with the critical exponent equal to the critical gap exponent. We also argued that thermodynamical  properties like the specific heat and the  magnetocaloric effect (instead of the usual internal energy and the magnetization) can detect the mentioned quantum entanglement in solid systems.  In addition, we suggest a tactic to find all critical temperatures, which is based on the derivative of the entanglement witness with respect to the temperature.

 \end{abstract}
\maketitle
\section{Introduction}\label{sec1}

The quantum entanglement is one of the most important prediction of modern quantum mechanics and  indeed a valuable resource in quantum information processing \cite {Wooters98, Bennett00, Amico08}. In fact entanglement is a unique quantum property of any nonlocal superposition-state of two or more quantum systems. Much effort is devoted to describing the nature of the entanglement \cite{Jafari1,Jafari2}.

A kind of innate entanglement so-called the thermal entanglement (TE) is of particular interest and demonstrates that non-local correlations persist even in the thermodynamic limit\cite{Nielsen98, Arnesen01}. 
It is believed that a connection between the quantum information theory and  condensed matter physics can be made by the study of TE,  zero-temperature entanglement and the relation between quantum phase transitions.
Since the thermal entanglement can be inferred by the macroscopic variables which these macroscopic variables are detected experimentally\cite{Ghosh03, Vedral03, Wiesniak05} many attempts have been dedicated to quantifying the TE.

One-dimensional spin-1/2 systems are a special and practical category in studying the thermal entanglement phenomenon\cite{Wang01-1, Wang02-1, Wang02-1-1, Zanardi02, Wang02-2, Kamta02, Wang02-3, Osborne02,  Yeo05, Amico07, Gong09, Werlang10, Maziero10, Ananikian12, Jafari3}. The ground state of the spin-1/2 antiferromagnetic XX chain is in the Luttinger liquid phase where the nearest neighbor spins are entangled. Increasing temperature, the TE is reduced and will be zero at a critical temperature ($T_{c}$) which is independent of a transverse magnetic field (TF). For values of the TF larger than the quantum critical field, there is no pairwise entanglement at zero temperature. In this case, adding temperature creates pairwise entanglement at a low-temperature interval\cite{Gong09}. In addition, the pairwise quantum discord is also studied recently\cite{Maziero10}. It is shown how quantum discord can be increased with temperature as the TF is varied.  The effect of a staggered magnetic field on the TE of the spin-1/2 XX model is also investigated\cite{Hide07} and it is found that the alternating magnetic field suppresses the TE.

Fundamentally, the magnetic behavior is determined by the Heisenberg model of interaction. In addition, the Dzyaloshinskii-Moriya (DM) interaction\cite{Dzyaloshinskii58, Moriya60} which is arising from the spin-orbit coupling, describes the superexchange  between the interacting spins and it is believed that it can  generate many surprising characteristics such as canting\cite{Coffey90} or the induced gap  in the 1D spin-1/2 isotropic Heisenberg model\cite{Affleck97}.  Some antiferromagnetic  systems are expected to be described by DM interaction such as $Cu(C_6D_5COO)_2 3D_2O$,\cite{Dender96, Dender97} $Yb_4As_3$,\cite{Kohgi01, Fulde95, Oshikawa99} $BaCu_2 Si_2O_7$,\cite{Tsukada01} $\alpha -Fe_2O_3$, $LaMnO_3$,\cite{Grande75}, $CuSe_{2}O_{5}$,\cite{Herak11, Herak13} $Cs_2 Cu Cl_4$,\cite{Povarov11} and $K_2V_3O_8$\cite{Greven95},  which exhibit unusual and interesting magnetic properties due to quantum fluctuations in the presence of an applied magnetic field.\cite{Grande75,Yildirim95,Katsumata01} $La_2CuO_4$ also belongs to the class of DM antiferromagnets, which is a parent compound of high- temperature superconductors.\cite{Kastner98} This has stimulated extensive investigations of the properties which are created from the DM interaction. On the other hand, for an explanation of the electric polarization behavior in multiferroic materials\cite{Cheong07}, an important and  sufficient  mechanism which is based on the DM interaction is proposed \cite{Sergienko06, Romhanyi11}.

The induced effects of the DM interaction on the TE is investigated only for two-qubit spin-1/2 XX chains\cite{Wang01-2, Zhang07, Zhang08, Albayrak09, Guo10}. It is believed that two or three qubit systems are not large enough to reveal
interesting correlation properties in condense matter physics. Also, it is possible that some of
the correlation phenomena to be  exclusively associated with the fact
that only two qubits are considered.  Moreover, as we have mentioned there are a large number of quasi-one-dimensional antiferromagnetic compounds which the low temperature behavior of them were studied experimentally. These compounds are very good candidate to study the effect of the DM interaction on the thermal entanglement. Therefore, in this paper we consider an infinite 1D spin-1/2 XX model with added DM interaction in a TF.  Using the Jordan-Wigner transformation we find an analytical solution for the TE between NN spins in the thermodynamic limit.  In the absence of the TF, in spite of the face that the DM interaction cannot make an impression in the amount of entanglement between NN spins at the zero temperature, but it can sufficiently affect it at the  finite temperature. In the presence of the TF, we show that depends on the value of the magnetic field, one or two critical temperatures can be found. In addition, an entanglement witness equivalent to the difference between the total energy ($U$) and the magnetic energy ($-h M$) is defined. Using entanglement witness, the parameter regions that entanglement can be detected in the solid state system are determined. It is also argued that the derivative of the witness with respect to the temperature, has good applicability to  detect the intermediate temperature interval where revival phenomenon is happened. Indeed, we suggest this technique to observe this phenomenon experimentally.

The paper is organized as follows. In the forthcoming section we
introduce the model and map it onto a pure 1D spin-1/2 XX model in a TF. In section III, we
present  our exact analytical results on the thermal behavior of the entanglement between NN spins. In section IV, we introduce an entanglement witness and explain how one can detect the quantum entanglement in the solid state systems. We conclude and summarize our results in section V.



\section{THE MODEL}\label{sec2}

We start our investigation with  the 1D spin-1/2 XX model with added DM interaction in a TF which the Hamiltonian is written as
\begin{eqnarray}
{\cal H}&=&J\sum_{j=1}^{N}(S_{j}^{x}~S_{j+1}^{x}+S_{j}^{y}~S_{j+1}^{y})+\overrightarrow{D}.\sum_{j=1}^{N}(\overrightarrow{S}_{j}\times \overrightarrow{S}_{j+1})  \nonumber\\
&-&h\sum_{j=1}^{N}S_{j}^{z}.
\label{Hamiltonian s}
\end{eqnarray}
 Where $S_{j}$ is the spin-1/2 operator on the $j$-th site, $J$ denotes the exchange coupling constant, $h$ is the TF and $\overrightarrow{D}$ is known as the DM vector. By considering uniform DM vector as $\overrightarrow{D}=D \hat{z}$, and doing the rotation\cite{Alcaraz90,Kaplan83,Jafari4,Jafari5} about the $z$ axis as $S_{j}^{\pm} \longrightarrow S_{j}^{\pm} \exp(\mp i\alpha)$ which  $\tan{\alpha}=\frac{-D}{J}$, the Hamiltonian is transformed to the following 1D spin-1/2 XX model in a TF

\begin{eqnarray}
{\cal
 H}=\tilde{J}\sum_{j=1}^{N}(\widetilde{S}_{j}^{x}\widetilde{S}_{j+1}^{x}+\widetilde{S}_{j}^{y}\widetilde{S}_{j+1}^{y})
  &-&h\sum_{j=1}^{N}\widetilde{S}_{j}^{z},
\label{Hamiltonian-tf}
\end{eqnarray}
 with an effective exchange $\tilde{J}=\sqrt{J^{2}+D^{2}}$. It is known, that at zero temperature, $T=0$, the ground state of the system is in the Luttinger liquid (LL) phase\cite{Takahashi99}. By increasing the TF from zero, up to the critical TF $h_c= \tilde{J}$, the ground state remains in the LL phase where a  quantum phase transition into the ferromagnetic phase with saturation magnetization along the TF will happen. At zero temperature, in the absence  of the TF, NN are entangled and by increasing the TF the concurrence decreases and will be equal to zero at the critical TF $h_{c}$.  In the saturated ferromagnetic phase, spins clearly are not entangled.

Theoretically, the energy spectrum is needed to investigate the thermodynamic properties of the model. In this respect, we implement the Jordan-Wigner transformation to fermionize
the transformed model (Eq.~(\ref{Hamiltonian-tf})). Using the Jordan-Wigner transformation

\begin{eqnarray}
\widetilde{S}_{j}^{z}&=&a_{j}^{\dagger}a_{j}-\frac{1}{2}, \\ \nonumber
\widetilde{S}_{j}^{+}&=&a_{j}^{\dagger} \exp(i\pi\sum_{l<j}a_{l}^{\dagger}a_{l}),\\ \nonumber
\widetilde{S}_{j}^{-}&=&a_{j} \exp(-i\pi\sum_{l<j}a_{l}^{\dagger}a_{l}),
\label{Hamiltonian}
\end{eqnarray}
 the transformed Hamiltonian is mapped onto a 1D model of noninteracting spinless fermions
\begin{eqnarray}
H_{f}= \frac{N h}{2}+\widetilde{J} \sum_{j} (a^{\dag}_{j}a_{j+1}+a^{\dag}_{j+1}a_{j})-h \sum_{j} a^{\dag}_{j}a_{j+1}. \end{eqnarray}
By performing a Fourier transformation into the momentum space as $a_{j} = \frac{1}{\sqrt{N}} \sum ^{N} _{j=1} e^{-ikj} a_{k}$, the diagonalized Hamiltonian is given by
\begin{eqnarray}
{\cal H}=\sum_{k=-\pi}^{\pi}\varepsilon(k) a_{k}^{\dagger}a_{k}.
\label{Hamiltonian d}
\end{eqnarray}
where $\varepsilon(k)$ is the dispersion relation
\begin{eqnarray}
\varepsilon(k) &=&  \widetilde{J} \cos(k)-h.
\end{eqnarray}

\section{Thermal entanglement}\label{sec4}

 We confine our interest to the entanglement between two sites which is measured by the concurrence. The concurrence between two spins at sites $i$ and $j$ in the ground state and at a finite temperature can be achieved from the corresponding reduced density matrix $\rho_{i,j}$, which in the standard basis $(\mid 11\rangle, \mid 10\rangle,\mid 01\rangle,\mid 00\rangle)$ can be expressed as\cite{Gong09}:

\[
\rho_{i,j} =
\left( {\begin{array}{cccc}
 \langle P_{i}^{\uparrow}P_{j}^{\uparrow}\rangle  & \langle P_{i}^{\uparrow} \sigma_{j}^{-}\rangle &  \langle \sigma_{i}^{-} P_{j}^{\uparrow} \rangle & \langle \sigma_{i}^{-} \sigma_{j}^{-}\rangle\\
 \langle P_{i}^{\uparrow} \sigma_{j}^{+}\rangle &  \langle P_{i}^{\uparrow}P_{j}^{\downarrow}\rangle  & \langle \sigma_{i}^{-} \sigma_{j}^{+}\rangle & \langle \sigma_{i}^{-} P_{j}^{\downarrow} \rangle\\
 \langle \sigma_{i}^{+} P_{j}^{\uparrow} \rangle & \langle \sigma_{i}^{+} \sigma_{j}^{-}\rangle & \langle P_{i}^{\downarrow}P_{j}^{\uparrow}\rangle  & \langle P_{i}^{\downarrow} \sigma_{j}^{-}\rangle\\
 \langle \sigma_{i}^{+} \sigma_{j}^{+}\rangle & \langle \sigma_{i}^{+} P_{j}^{\downarrow} \rangle & \langle P_{i}^{\downarrow} \sigma_{j}^{+}\rangle & \langle P_{i}^{\downarrow}P_{j}^{\downarrow}\rangle
 \end{array} } \right),
\]
where $P^{\uparrow}=\frac{1}{2}(1+\sigma^{z}), P^{\downarrow}=\frac{1}{2}(1-\sigma^{z})$. The brackets symbolize the ground-state and thermodynamic average values at zero and finite temperature, respectively and  $\sigma^{x}, \sigma^{y},\sigma^{z}$ are Pauli matrices\cite {Gong09}. The concurrence between two spins is given through  $C_{j}=max(0,\lambda_{1}-  \lambda_{2} -\lambda_{3} -\lambda_{4}$) where  $\lambda_{i}$ is the square root of the eigenvalue of   $R=\rho_{j,j+1}  \tilde{\rho}_{j,j+1}$ and  $\tilde{\rho}_{j,j+1}=(\sigma_{j}^{y}\otimes\sigma_{j+1}^{y})~\rho^{\ast}(\sigma_{j}^{y}\otimes\sigma_{j+1}^{y})$. By applying the Jordan-Wigner transformation, the reduced density matrix will be written as

\[
\rho_{i,j} =
\left( {\begin{array}{cccc}
  X_{j}^{+}  & 0 &  0 & 0\\
    0 &  Y_{j}^{+}  & Z_{j}^{*} & 0\\
 0 & Z_{j} & Y_{j}^{-}  & 0\\
 0 & 0 & 0 & X_{j}^{-}
 \end{array} } \right),
\]
where $X_{j}^{+}=\langle n_j n_{j+1}\rangle (n_j=a_{j}^{\dagger}a_{j})$,  $Y_{j}^{+}=\langle n_j(1-n_{j+1})\rangle$, $Y_{j}^{-}=\langle n_{j+1}(1-n_{j})\rangle$, $Z_j=\langle a_{j}^{\dagger}a_{j+1} \rangle$ and $X_{j}^{-}=\langle 1-n_j- n_{j+1}+n_j n_{j+1}\rangle$. Thus the concurrence is transformed into

\begin{eqnarray}
C_{j}=max\{0,2 (|Z_{j}|-\sqrt{X_{j}^{+} X_{j}^{-}})\},
\label{concurrence}
\end{eqnarray}
where
\begin{eqnarray}
Z_{j}=\frac{1}{2 \pi} \int_{-\pi}^{\pi} \frac{e^{i k}}{1+e^{\beta \varepsilon(k)}} dk,
\end{eqnarray}
\begin{eqnarray}
n_{j}=\frac{1}{2 \pi} \int_{-\pi}^{\pi} \frac{1}{1+e^{\beta \varepsilon(k)}} dk,
\end{eqnarray}
where $\beta=\frac{1}{k_{B} T}$ and the Boltzmann constant is taken as $k_B=1$. One should note that the Fermi distribution function is $f(k)=\frac{1}{1+e^{\beta \varepsilon(k)}}$.   Using the solution of the retarded Green's function\cite{Fetter71},  $X_{j}^{+}$ approximately is obtained as $X_{j}^{+}=\langle n_j \rangle^{2}-Z_{j}^{2}$.

\begin{figure}
\centerline{\psfig{file=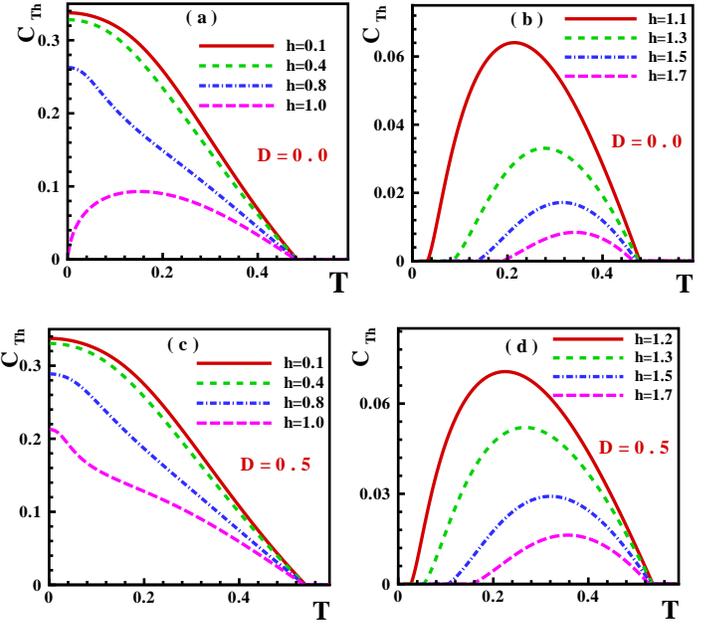,width=3.65in}}
 \caption{(Color online.) The thermal entanglement between NN spins as a function of the temperature for values of the TF less than quantum critical point (a) $D=0$ (c) $D=0.5$, and larger than the quantum critical point (b) $D=0$ (d) $D=0.5$.   }\label{Ent-temperature}
\end{figure}

 In  following, we investigate the concurrence between NN spins for different values of ${J}$, ${D}$ and ${h}$ and depict the behavior of  concurrence  with respect to each of the above parameters. The thermal behavior of the concurrence between NN spins in the pure spin=1/2 Heisenberg XX model ($D=0$) in a TF, has been studied\cite{Gong09}. It was found that the TE reduces by increasing temperature and will be zero at a critical temperature which was shown to be independent of the TF. On the other hand, for the values of the TF which are more than the quantum critical TF, the amount of concurrence will be retrieved so revival phenomenon can happen for these values of the TF.

Now, we try to depict a physical picture of the DM effect on the thermal behavior of the concurrence in the introduced model.   In Fig.~\ref{Ent-temperature}, we have presented our analytical results on the behavior of the TE of the model in the absence of the DM interaction ((a) and (b)) and in the presence of the DM interaction ((c) and (d)). It can be  clearly seen, for values of the TF  less than quantum critical point (Fig.~\ref{Ent-temperature}(a) and (c)),  the TE decreases with increasing the temperature and vanishes at a field-independent critical temperature ($T_{c}$). From the physical point of view, the number of excited states
involved depends on temperature, where more states are added as the temperature is raised. This mixing of excited
states with the ground state act as a destructive noise that reduces the amount of entanglement contained in the
 system. When the temperature reaches certain value, which varies based on the system characteristics and parameters
values, the amount of noise created by the excited states due to thermal fluctuations is sufficient to turn the system
 into a disentangled state. This temperature is known as the critical temperature, where below it the system is guaranteed to be entangled.
In principle, at this temperature all quantum correlations will be destroyed by classical thermal fluctuations. Therefore at $T=T_{c}$, $C_{Th}=0$ and one can derive\cite{Gong09}

\begin{eqnarray}
\langle n_{j} \rangle - \langle n_{j} \rangle^{2}=-\sqrt{2} Z_{j}-Z_{j}^{2}.
\end{eqnarray}
In the absence of the TF, $\langle n_{j} \rangle=1/2$ at any temperature. Thus the critical temperature will be find by solving the following equation
\begin{eqnarray}
\frac{\sqrt{2}-1}{2}=\frac{\widetilde{J}}{\pi T_{c}} \int_{0}^{1} \frac{\sqrt{1-x^{2}}}{1+\cosh(\widetilde{J} x/T_{c})} dx,~~~(x=\cos(k)),\nonumber \\
\end{eqnarray}
which indicates that the critical temperature in the absence of the TF is related to the DM interaction as
\begin{eqnarray}
T_{c}\simeq0.48 J \sqrt{1+D^2/J^2},
\label{TC}
\end{eqnarray}
Which is smaller than the critical temperature in the two-qubit systems\cite{Wang01-2, Zhang08, Albayrak09}.

On the other hand for values of the TF more than quantum critical point  (Fig.~\ref{Ent-temperature}(b) and (d)), NN spins are not entangled at $T=0$. By increasing the temperature from zero, NN spins remain unentangled up to the first critical  temperature $T_{c_{1}}(h)$. As soon as the temperature increases from $T_{c_{1}}$, the TE regains  and takes a maximum value and then decrease and reaches to zero at  the second critical temperature $T_{c_{2}}$. The existence of the second critical temperature is completely natural, since sufficiently large thermal fluctuations will destroy all classical and quantum correlations. It  is seen that the amount of $T_{c_{1}}$ increases when the external TF raises, but  $T_{c_{2}}$ is almost field-independent. Therefore,  the width  of the temperature interval which within the NN spins become entangled is getting smaller by increasing the TF. We have calculated numerically the width of this entangled region as a function of the $h-h_c$ and results show a linear scaling behavior as
\begin{eqnarray}
T_{c_{2}}-T_{c_{1}}=0.932-0.381\times(h-h_c),~~h\geq h_c.
\end{eqnarray}
 In addition, the maximum value of the entanglement in the mentioned temperature interval behaves as
\begin{eqnarray}
C_{Th}^{max}=0.334-0.017\times(h-h_c)^{2},~~h\geq h_c.
\end{eqnarray}
At the quantum critical TF $h=h_c$ and zero temperature, the system is at the quantum critical point and the entanglement is zero ($C_{th}=0$). There are many studies on the behavior of entanglement close to the quantum phase transition point. Recently, the study of the role of the temperature on the quantum properties of entanglement is regarded\cite{Amico07-1}. It is suggested that the entanglement sensitivity to thermal and to quantum fluctuations obeys universal finite temperature scaling laws. In following, we study the scaling behavior of the TE at a finite temperature neighborhood of $h_c$. One should note, though the TE is not diverging at the quantum critical point but it is affected by the quantum criticality. We analyzed our analytical results and found that as soon as the temperature increases from zero, the TE between NN spins increases from zero and shows a scaling behavior as
\begin{equation}
C_{th}\propto T^{\varepsilon},
\label{tr-magnetization}
\end{equation}
with the critical exponent $\varepsilon=0.70\pm 0.04$. It is surprising that the mentioned critical exponent is almost the same as the critical exponent of the energy gap ($\varepsilon=2/3$)\cite{Fouet04, Mahdavifar07} in the vicinity of this critical TF.

\begin{figure}[t]
\centerline{\psfig{file=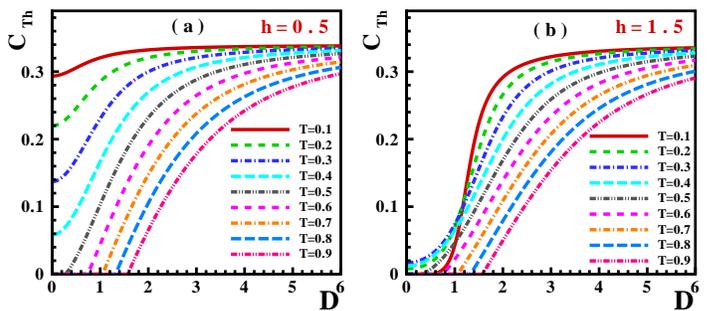,width=3.65in}}
 \caption{(Color online.) The thermal entanglement between NN spins as a function of the DM interaction for different values of temperature and the TF (a) $h=0.5$ (b) $h=1.5$. }\label{Concurrence-h}
\end{figure}

To have a deep insight into the nature of the system, we continue our study through the case of fixed TF. In this case one can find the induced effects of the DM interaction on the quantum correlations of the spin-1/2 Heisenberg XX model at finite temperature. We have presented our results in Fig.~\ref{Concurrence-h} for the exchange $J=1$ and different values of the temperature. As it is seen from Fig.~\ref{Concurrence-h}, increase of the DM interaction leads to an increase in  the amount of the TE till it reaches to its saturation value ($\simeq1/3$).  By the way, at low temperatures, this saturation value can be achieved at small values of the DM interaction while, as the temperature increases reaching to this saturation value happens at larger values of the DM interaction due to the classical thermal fluctuations. The increasing of the TE with DM interaction at fixed temperature is a consequence of the fact that, when the DM interaction is turned on, the low-lying excited states tend to be more correlated. Important is that the increasing behavior of the TE in respect to the DM interaction at low temperatures is field-independent.

\section{Entanglement Witness}\label{sec2}

In recent years, the realization that entanglement can also affect macroscopic properties of bulk solid-state systems, has increased the interest in characterizations of entanglement in terms of macroscopic thermodynamical\cite{Ghosh03, Vedral03, Brukner04} observables. The entanglement witness is called an observation  which can distinguish between
entangled and separable states in the quantum physics\cite{Terhal00}. In principle, entanglement witness has positive expectation value for separable states and a negative one for some specific, entangled states. From an experimental point of view, several methods for detection of entanglement using witness operators have been proposed \cite{Boure04}. As a result of these studies, entanglement witnesses have been obtained in terms of expectation values of thermodynamical observables such as internal energy, magnetization and magnetic susceptibility.

\begin{figure}
\centerline{\psfig{file=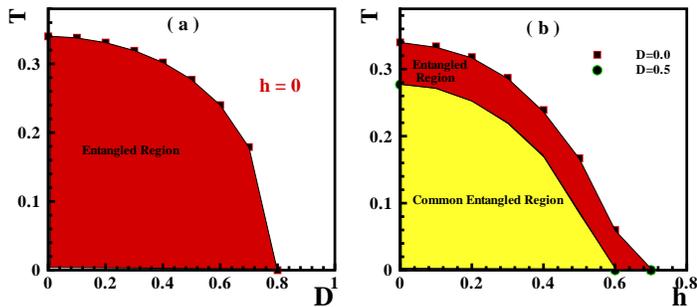,width=3.65in}}
 \caption{(Color online.) The parameter region of temperature $T$ and (a) the DM interaction ($h=0$), (b) the magnetic field ($D=0, 0.5$). The thermodynamic witness $\frac{|U+hM|}{N \widetilde{J}}$ is more than $0.25$ at temperatures less than critical line which detects entanglement in the system.  }\label{witness-o}
\end{figure}
Here, we define the entanglement witness as\cite{Toth05, Hide07}
\begin{eqnarray}
W=\frac{1}{\beta N} \frac{\partial \ln Z}{\partial \widetilde{J}}&=&\frac{1}{N}\sum_{j=1}^{N}(\langle \widetilde{S}_{j}^{x}\widetilde{S}_{j+1}^{x}\rangle+\langle \widetilde{S}_{j}^{y}\widetilde{S}_{j+1}^{y}\rangle)\nonumber \\
&=&\frac{U+hM}{N \widetilde{J}},
\end{eqnarray}
where $U=\langle H \rangle$  and $M=\sum_{j=1} \widetilde{S}_{j}^{z}$ are the total energy and the magnetization respectively. Our witness is physically equivalent to the difference between the total energy ($U$) and the magnetic energy ($-h M$). If,
\begin{eqnarray}
\frac{| U+hM |}{N \widetilde{J}}>0.25,
\end{eqnarray}
then the system is in an entangled state. In the absence of the magnetic field, the magnetization is zero and the thermodynamic witness reduces to $\frac{|U|}{N \widetilde{J}}>0.25$. In this case the concurrence is given by $max \{0, \frac{|U|}{N \widetilde{J}}-0.25\}$.

Applying the fermionized operators, the entanglement witness is obtained as
\begin{eqnarray}
W=|\frac{1}{2 \pi} \int_{-\pi}^{\pi} \frac{\cos k}{1+e^{\beta \varepsilon(k)}} dk|.
\label{witness}
\end{eqnarray}
Using this equation we have determined the parameter regions where entanglement can be detected in the solid state systems. Results are presented in Fig.~\ref{witness-o} (a) and (b). In the absence of the magnetic field (Fig.~\ref{witness-o}(a)), by adding DM interaction, the critical temperature decreases and in the region $D\succeq 0.8$ the detection of the entanglement in the spin-1/2 XX Heisenberg solid state system from entanglement witness $\frac{|U+hM|}{N \widetilde{J}}$ is impossible. The effect of the TF on the critical temperature is shown in Fig.~\ref{witness-o}(b). In the absence of the DM interaction, critical temperature decreases with increasing the TF and can not determine experimentally for values of the TF $h\succeq 0.7$ in complete agreement with the result of Ref.~[\onlinecite{Brukner04}]. In addition, applying the DM interaction, critical temperature decreases and for example in the region of TF, $h(D=0.5)\succeq 0.6$, is impossible to detect experimentally by focussing on witness.

Comparing results of TE (Eq.~(\ref{concurrence})) and witness (Eq.~(\ref{witness})), some disagreement is seen. Firstly, equation ~(\ref{TC}) shows that the critical temperature in the absence of the TF should be increased with increasing the DM interaction, which was not observed by measuring the thermodynamic witness. Secondly, the TE shows that the system will be entangled in an intermediate region of temperature ($T_{c_1}<T<T_{c_2}$) for fields larger than the quantum critical point, which the detection of this phenomenon is impossible by measuring the thermodynamic witness. In following, we propose a tactic to resolve  the mentioned disagreement.

\begin{figure}
\centerline{\psfig{file=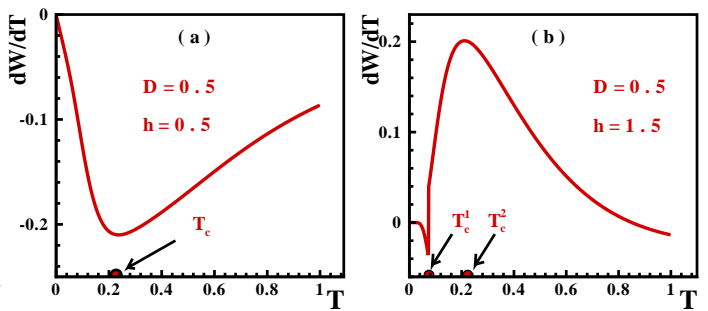,width=3.65in}}
 \caption{(Color online.) The derivative of our witness with respect to the temperature, $dW/dT$, as a function of temperature for DM interaction $D=0.5$ and different values of the TF (a) $h=0.5<h_c$, (b) $h=1.5>h_c$.   }\label{D-witness}
\end{figure}

In the topic of the critical phenomena, it is known that the derivative of the response functions with respect to the control parameter can provide very useful and interesting results about the critical points.
 Inspired with this subject,  instead of the witness we focus on the derivative of the witness with respect to the temperature
\begin{eqnarray}
\frac{dW}{dT}&=&\frac{\partial}{\partial T}\{|\frac{1}{N}\sum_{j=1}^{N}(\langle \widetilde{S}_{j}^{x}\widetilde{S}_{j+1}^{x}\rangle+\langle \widetilde{S}_{j}^{y}\widetilde{S}_{j+1}^{y}\rangle)|\}\nonumber \\
&=&\frac{\partial}{\partial T}(\frac{|U+hM|}{N \widetilde{J}}).
\end{eqnarray}
We know that the specific heat is defined as
\begin{eqnarray}
C_v=\frac{\partial U}{\partial T},
\end{eqnarray}
and on the other hand the derivative of the magnetization with respect to the temperature is known as the magnetocaloric effect
\begin{eqnarray}
-(\frac{\partial M}{\partial T})|_h=(\delta Q/\delta h)/T,
\end{eqnarray}
where $\delta Q$ is the amount of heat created or absorbed by the solid state sample for a field change $\delta h$ due to the magnetocaloric effect. Thus, the derivative of our witness with respect to the temperature is physically equivalent to the difference between the specific heat ($C_v$) and the magnetocaloric effect ($-(\frac{\partial M}{\partial T})|_{h}$).

\begin{figure}
\centerline{\psfig{file=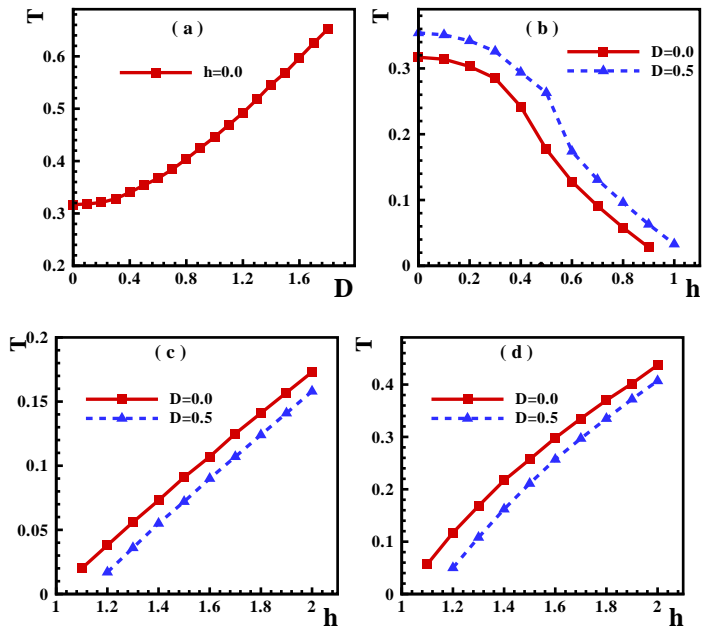,width=3.65in}}
 \caption{(Color online.) The parameter region of first critical temperature $T$ and (a) the DM interaction ($h=0$), (b) the TF ($D=0, 0.5$) . Figures (c) and (d) show the results on the first and second critical temperatures for values of the TF larger that the quantum critical field.    }\label{T-witness}
\end{figure}

Fig.~\ref{D-witness} shows the $dW/dT$ as a function of temperature for DM interaction $D=0.5$ and different values of the TF (a) $h=0.5<h_{c}$ and (b) $h=1.5>h_{c}$. As it is seen, the derivative of the witness with respect to the temperature shows one or two extremum at certain temperatures for values of the TF less or more than the quantum critical field. The same results are found for other values of the DM interaction. As a new approach, we suggest these certain temperatures as the critical temperatures to observe the entanglement in the solid state systems.

Using the mentioned approach, we have calculated the critical temperatures and results are presented in Fig.~\ref{T-witness}. It is clearly seen from Fig.~\ref{T-witness}(a) that, in the absence of the TF, the critical temperature at which the system is entangled below it, increases by the increase of the DM interaction in complete agreement with TE (Eq.~(\ref{TC})). The effect of the TF is studied in Fig.~\ref{T-witness}(b), (c) and (d). From Fig.~\ref{T-witness}(b) is seen that the critical temperature decreases with increasing the TF and vanishes exactly at the quantum critical field. The value of the critical temperature increases by increasing the DM interaction in the presence of the TF, which is also in agreement with the TE. As soon as the TF becomes larger than the quantum critical point, the first ($T_{c_{1}}$) and second ($T_{c_{1}}$) critical temperatures will be observed. The first critical temperature $T_{c_{1}}$, increases by increasing the TF (Fig.~\ref{T-witness}(c)) in complete agreement with the TE. But, the second one (Fig.~\ref{T-witness}(d)) also increases by TF which is not in agreement with the TE. Anyway, we believe that the increasing behavior must be related to increasing of the energy gap in the saturated ferromagnetic phase.


\section{Conclusion}\label{sec2}

We considered the 1D spin-1/2 XX model with added Dzyaloshinskii-Moriya interaction in a TF. At zero temperature, it has been found that the NN spins are entangled in the absence of DM interaction and  TF. Adding the DM interaction does not affect the  amount of entanglement between NN spins. But, since the TF causes a quantum phase transition into a saturated ferromagnetic phase, thus the entanglement between NN spins decreases with  increasing  the TF and will be zero at the  quantum critical field $h=h_{c}(D)$.

In this work, we studied the temperature dependence of the entanglement between NN spins using the fermionization technique. It is found that the TE in the region $h<h_c(D)$ decreases by thermal fluctuations and will be zero at a field-independent critical temperature $T=T_c (D)$. Which shows that at this critical temperature all quantum correlations will be destroyed by classical thermal fluctuations. The critical temperature $T_c$ increases by increasing the DM interaction.  It is inferred that in  the absence of the TF, in spite of the face that the DM interaction cannot make an impression in the amount of entanglement between NN spins at zero temperature, but it can sufficiently affect it at the  finite temperature.

At the critical field $h=h_c(D)$ and zero temperature, the system is at the quantum critical point and the entanglement is zero. The scaling behavior of the TE in a finite temperature neighborhood of $h_c$ is studied by analyzing our exact results. It is found that as soon as the temperature increases from zero, the TE increases and shows a scaling behavior as $C_{th}\propto T^{\varepsilon}$ with the critical exponent $\varepsilon=0.70\pm 0.04$ the same as the critical exponent of the energy gap in the vicinity of this critical field.

  For values of the TF more than quantum critical point $h>h_c(D)$, by the increase of the temperature from zero, NN spins remain unentangled up to a first critical temperature $T_{c_{1}}$. As soon as the temperature increases from $T_{c_{1}}$, the thermal entanglement regains  and takes a maximum value and then decreases and reaches to zero at  the second critical temperature $T_{c_{2}}$. The existence of the second critical temperature is completely natural, since sufficiently large thermal fluctuations will destroy all classical and quantum correlations.

Finally, an entanglement witness equivalent to the difference between the total energy ($U$) and the magnetic energy ($-h M$) is defined. Using entanglement witness, the parameter regions that entanglement can be detected in the solid state system are determined. In the regions of $D(h=0)\succeq 0.8$ and $h(D=0)\succeq 0.7$, it is impossible to detect the entanglement in the XX Heisenberg solid state system. Generally, applying the DM interaction reduces the area of the entangled region of the spin-1/2 XX model in the TF. By comparing results of the TE and our witness, some disagreement are presented. First, equation ~(\ref{TC}) shows that the critical temperature in the absence of the magnetic field should be increased by increasing the DM interaction, which does not observe by measuring the thermodynamic witness. Second, TE shows that the system will be entangled in an intermediate region of temperature ($T_{c_1}<T<T_{c_2}$) for fields larger than the quantum critical point, which the detection of this phenomenon is impossible by measuring the thermodynamic witness. To resolve these disagreement,  we have suggested to focus on the derivative of the witness with respect to the temperature which is physically equivalent to the difference between the specific heat ($C_v$) and the magnetocaloric effect ($-(\frac{\partial M}{\partial T})|_{h}$). However, we suggest that it is possible to detect all critical temperatures from experimental point of view if one focuses on the derivative of the witness with respect to the temperature.


\vspace{0.3cm}


\end{document}